\documentclass{ws-ijmpd}
\usepackage{amssymb}

\begin{document}

\def\nocropmarks{\vskip5pt\phantom{cropmarks}}
\let\trimmarks\nocropmarks
\markboth{Remo Ruffini and Luca Vitagliano}
{Energy Extraction From Gravitational Collapse to Static Black Holes}

\catchline{}{}{}

\title{ENERGY EXTRACTION FROM GRAVITATIONAL COLLAPSE TO STATIC BLACK HOLES}

\author{\footnotesize REMO RUFFINI\footnote{
E-mail: ruffini@icra.it} and LUCA VITAGLIANO\footnote{E-mail: vitagliano@icra.it}}

\address{International Centre for Relativistic Astrophysics, Department of Physics,
Rome University ``La Sapienza", P.le Aldo Moro 5, 00185 Rome, Italy}

\maketitle

\pub{Received (received date)}{Revised (revised date)}

\begin{abstract}
The mass--energy formula of black holes implies that up to 50\% of the energy can be extracted from a static black hole. Such a result is reexamined using the recently established analytic formulas for the collapse of a shell and expression for the irreducible mass of a static black hole. It is shown that the efficiency of energy extraction process during the formation of the black hole is linked in an essential way to the gravitational binding energy, the formation of the horizon and the reduction
of the kinetic energy of implosion. Here a maximum efficiency of 50\% in the extraction of the mass energy is shown to be generally attainable in the collapse of a spherically symmetric shell: surprisingly this result holds as well in the two limiting cases of the Schwarzschild and extreme Reissner-Nordstr\"{o}m space-times. Moreover, the analytic expression recently found for the implosion of a spherical shell onto an already formed black hole leads to a new exact analytic expression for the energy extraction which results in an efficiency strictly less than 100\% for any physical implementable
process. There appears to be no incompatibility between General Relativity and Thermodynamics at this classical level.
\end{abstract}

\section*{  }
The mass--energy formula for a black hole endowed with electromagnetic structure
(EMBH) (see Christodoulou, Ruffini (1971)\cite{CR71}) has opened the field of the energetics of black holes (see Ruffini, 1973\cite{rlh}) which is at the very heart of the explanation of GRBs (Ruffini, 1978\cite{r78}, Ruffini, et al., 2003\cite{r03}). The recent formulation of the analytic solution for a self gravitating charged shell\cite{CRV02} has led to a new formula, expressed in Ruffini, Vitagliano (2002)\cite{RV02}, relating the irreducible
mass $M_{\mathrm{irr}}$ of the final black hole to the gravitational binding
energy and the kinetic energy of implosion evaluated at the horizon. The aim of this article is to point out how this new formula for the $M_{\mathrm{irr}}$ leads to a deeper physical understanding of the role of the gravitational interaction in the maximum energy extraction process of an EMBH. This formula can also be of assistance in clarifying some long lasting epistemological issue on the role of general relativity, quantum theory and thermodynamics.

It is
well known that if a spherically symmetric mass distribution without any
electromagnetic structure undergoes free gravitational collapse, its total
mass-energy $M$ is conserved according to the Birkhoff theorem: the increase
in the kinetic energy of implosion is balanced by the increase in the
gravitational energy of the system. If one considers the possibility that part
of the kinetic energy of implosion is extracted then the situation is very
different: configurations of smaller mass-energy and greater density can be
attained without violating Birkhoff theorem.

We illustrate our considerations with two examples: one has found confirmation
from astrophysical observations, the other promises to be of relevance for
gamma ray bursts (GRBs) (see Ruffini, Vitagliano (2002)\cite{RV02}). Concerning the first example, it is
well known from the work of Landau\cite{L32} that at the endpoint of
thermonuclear evolution, the gravitational collapse of a spherically symmetric
star can be stopped by the Fermi pressure of the degenerate electron gas
(white dwarf). A configuration of equilibrium can be found all the way up to
the critical number of particles
\begin{equation}
N_{\mathrm{crit}}=0.775\tfrac{m_{Pl}^{3}}{m_{0}^{3}},
\end{equation}
where the factor $0.775$ comes from the coefficient $\tfrac{3.098}{\mu^{2}}$
of the solution of the Lane-Emden equation with polytropic index $n=3$, and
$m_{Pl}=\sqrt{\tfrac{\hbar c}{G}}$ is the Planck mass, $m_{0}$ is the nucleon
mass and $\mu$ the average number of electrons per nucleon. As the kinetic
energy of implosion is carried away by radiation the star settles down to a
configuration of mass
\begin{equation}
M=N_{\mathrm{crit}}m_{0}-U, \label{BE}%
\end{equation}
where the gravitational binding energy $U$ can be as high as $5.72\times
10^{-4}N_{\mathrm{crit}}m_{0}$.

Similarly Gamov\cite{G51} has shown that a gravitational collapse process to
still higher densities can be stopped by the Fermi pressure of the neutrons
(neutron star) and Oppenheimer\cite{OV39} has shown that, if the effects of
strong interactions are neglected, a configuration of equilibrium exists also
in this case all the way up to a critical number of particles
\begin{equation}
N_{\mathrm{crit}}=0.398\tfrac{m_{Pl}^{3}}{m_{0}^{3}},
\end{equation}
where the factor $0.398$ comes now from the integration of the
Tolman-Oppenheimer-Volkoff equation (see e.g. Harrison et al. (1965)\cite{HTWW}). If the kinetic
energy of implosion is again carried away by radiation of photons or neutrinos
and antineutrinos the final configuration is characterized by the formula
(\ref{BE}) with $U\lesssim2.48\times10^{-2}N_{\mathrm{crit}}m_{0}$. These
considerations and the existence of such large values of the gravitational
binding energy have been at the heart of the explanation of astrophysical
phenomena such as red-giant stars and supernovae: the corresponding
measurements of the masses of neutron stars and white dwarfs have been carried
out with unprecedented accuracy in binary systems.\cite{GR75}

From a theoretical physics point of view it is still an open question how far
such a sequence can go: using causality nonviolating interactions, can one
find a sequence of braking and energy extraction processes by which the
density and the gravitational binding energy can increase indefinitely and the
mass-energy of the collapsed object be reduced at will? This question can also
be formulated in the mass-formula language of a black hole given in
Christodoulou, Ruffini (1971)\cite{CR71} (see also Ruffini, Vitagliano (2002)\cite{RV02}): given a collapsing core of nucleons with a
given rest mass-energy $M_{0}$, what is the minimum irreducible mass of the
black hole which is formed?

Following Cherubini, Ruffini, Vitagliano (2002)\cite{CRV02} and Ruffini, Vitagliano (2002)\cite{RV02}, consider a spherical shell of rest mass
$M_{0}$ collapsing in a flat space-time. In the neutral case the irreducible
mass of the final black hole satisfies the equation (see Ruffini, Vitagliano (2002)\cite{RV02})
\begin{equation}
M_{\mathrm{irr}}=M=M_{0}-\tfrac{M_{0}^{2}}{2r_{+}}+T_{+}, \label{Mirr2}%
\end{equation}
where $M$ is the total energy of the collapsing shell and $T_{+}$ the kinetic
energy at the horizon $r_{+}$. Recall that the area $S$ of the horizon is\cite{CR71}
\begin{equation}
S=4\pi r_{+}^{2}=16\pi M_{\mathrm{irr}}^{2} \label{S}%
\end{equation}
where $r_{+}=2M_{\mathrm{irr}}$ is the horizon radius. The minimum irreducible
mass $M_{\mathrm{irr}}^{\left(  {\mathrm{min}}\right)  }$ is obtained when the
kinetic energy at the horizon $T_{+}$ is $0$, that is when the entire kinetic
energy $T_{+}$ has been extracted. We then obtain the simple result
\begin{equation}
M_{\mathrm{irr}}^{\left(  \mathrm{min}\right)  }=\tfrac{M_{0}}{2}.
\label{Mirrmin}%
\end{equation}
We conclude that in the gravitational collapse of a spherical shell of rest
mass $M_{0}$ at rest at infinity (initial energy $M_{\mathrm{i}}=M_{0}$), an
energy up to $50\%$ of $M_{0}c^{2}$ can in principle be extracted, by braking
processes of the kinetic energy. In this limiting case the shell crosses the
horizon with $T_{+}=0$. The limit $\tfrac{M_{0}}{2}$ in the extractable
kinetic energy can further increase if the collapsing shell is endowed with
kinetic energy at infinity, since all that kinetic energy is in principle extractable.

\begin{figure}[ptb]
\begin{center}
\includegraphics[
height=3.5423in,
width=5.0574in
]{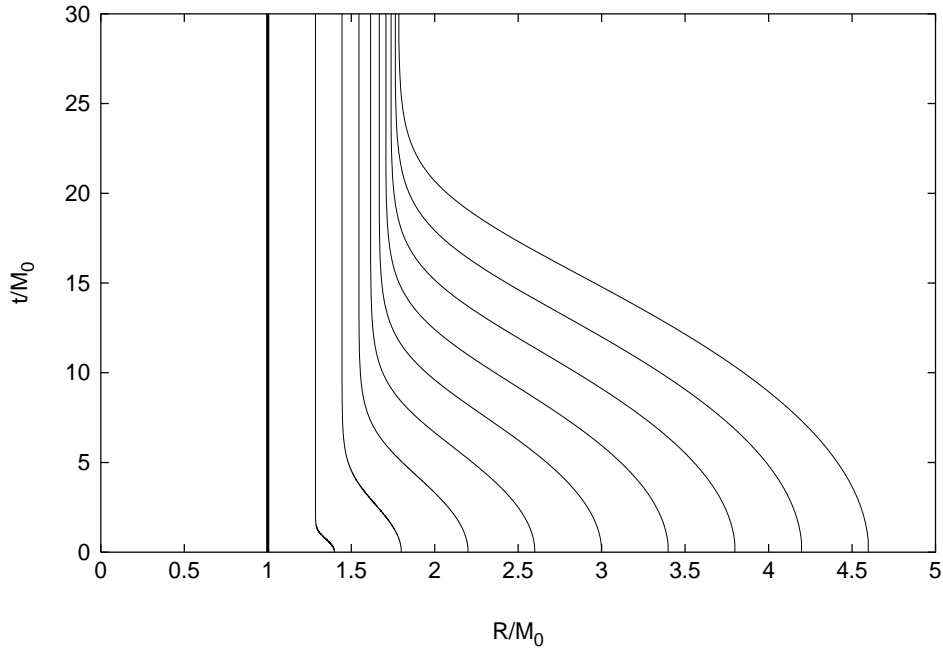}
\end{center}
\caption{Collapse curves for neutral shells with rest mass $M_{0}$ starting at rest at selected radii 
$R^{\ast}$ computed by using the exact solutions given in Cherubini, Ruffini, Vitagliano 
(2002).$^{2}$ A different value of $M_{\mathrm{irr}}$ (and therefore of $r_{+}$) 
corresponds to each curve. The time parameter is the Schwarzschild time coordinate $t$ and 
the asymptotic behaviour at the respective horizons is evident. The limiting configuration 
$M_{\mathrm{irr}}=\tfrac{M_{0}}{2}$ (solid line) corresponds to the case in which the 
shell is trapped, at the very beginning of its motion, by the formation of the horizon.}
\label{fig1}
\end{figure}

\begin{figure}[ptb]
\begin{center}
\includegraphics[height=11.21cm]{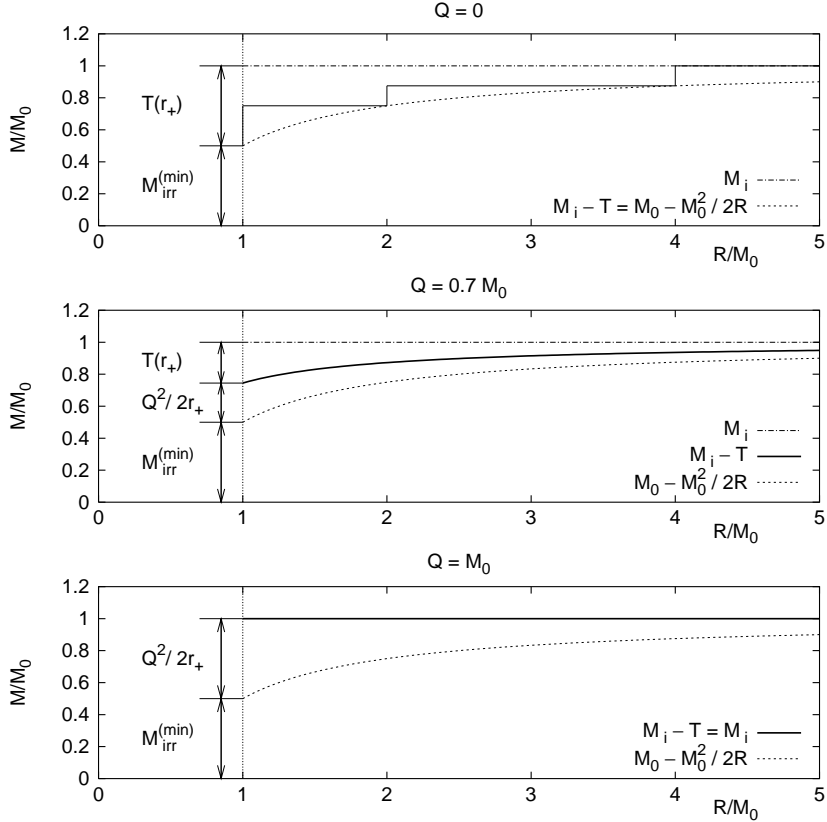}
\end{center}
\caption{Energetics of a shell such that $M_{\mathrm{i}}=M_{0}
$, for selected values of the charge. In the first diagram
$Q=0$; the dashed line represents the total energy for a
gravitational collapse without any braking process as a function of the
radius $R$ of the shell; the solid, stepwise line represents a
collapse with suitable braking of the kinetc energy of implosion at selected
radii; the dotted line represents the rest mass energy plus the gravitational
binding energy. In the second and third diagram $Q/M_{0}=0.7$,
$Q/M_{0}=1$ respectively; the dashed and the dotted lines have
the same meaning as above; the solid lines represent the total energy minus
the kinetic energy. The region between the solid line and the dotted line
corresponds to the stored electromagnetic energy. The region between the
dashed line and the solid line corresponds to the kinetic energy of collapse.
In all the cases the sum of the kinetic energy and the electromagnetic energy
at the horizon is 50\% of $M_{0}$. Both the electromagnetic and
the kinetic energy are extractable. It is most remarkable that the same
underlying process occurs in the three cases: the role of the electromagnetic
interaction is twofold: a) to reduce the kinetic energy of implosion by the
Coulomb repulsion of the shell; b) to store such an energy in the region
around the EMBH. The stored electromagnetic energy is extractable as shown in
Ruffini, Vitagliano (2002).$^{3}$}
\label{fig3}%
\end{figure}

In order to illustrate the physical reasons for this result, using the
formulas of Cherubini, Ruffini, Vitagliano (2002)\cite{CRV02}, we have represented in Fig.~\ref{fig1} the world lines
of spherical shells of the same rest mass $M_{0}$, starting their
gravitational collapse at rest at selected radii $R^{\ast}$. These initial
conditions can be implemented by performing suitable braking of the collapsing
shell and concurrent kinetic energy extraction processes at progressively
smaller radii (see also Fig.~\ref{fig3}). The reason for the existence of the
minimum (\ref{Mirrmin}) in the black hole mass is the ``self closure''
occurring by the formation of a horizon in the initial configuration (thick
line in Fig.~\ref{fig1}).

Is the limit $M_{\mathrm{irr}}\rightarrow\tfrac{M_{0}}{2}$ actually attainable
without violating causality? Let us consider a collapsing shell with charge
$Q$. If $M\geq Q$ an EMBH is formed. As pointed out in Ruffini, Vitagliano (2002)\cite{RV02} the
irreducible mass of the final EMBH does not depend on the charge $Q$.
Therefore Eqs.~(\ref{Mirr2}) and (\ref{Mirrmin}) still hold in the charged
case with $r_{+}=M+\sqrt{M^{2}-Q^{2}}$. In Fig.~\ref{fig3} we consider the
special case in which the shell is initially at rest at infinity, i.e. has
initial energy $M_{\mathrm{i}}=M_{0}$, for three different values of the
charge $Q$. We plot the initial energy $M_{i}$, the energy of the system when
all the kinetic energy of implosion has been extracted as well as the sum of
the rest mass energy and the gravitational binding energy $-\tfrac{M_{0}^{2}%
}{2R}$ of the system (here $R$ is the radius of the shell). In the extreme
case $Q=M_{0}$, the shell is in equilibrium at all radii (see Cherubini, Ruffini, Vitagliano (2002)\cite{CRV02}) and
the kinetic energy is identically zero. In all three cases, the sum of the
extractable kinetic energy $T$ and the electromagnetic energy $\tfrac{Q^{2}%
}{2R}$ reaches $50\%$ of the rest mass energy at the horizon, according to Eq.~(\ref{Mirrmin}).

What is the role of the electromagnetic field here? If we consider the case of
a charged shell with $Q\simeq M_{0}$, the electromagnetic repulsion implements
the braking process and the extractable energy is entirely stored in the
electromagnetic field surrounding the EMBH (see Ruffini, Vitagliano (2002)\cite{RV02}). In Ruffini, Vitagliano (2002)\cite{RV02} we have
outlined two different processes of electromagnetic energy extraction. We
emphasize here that the extraction of $50\%$ of the mass-energy of an EMBH is
not specifically linked to the electromagnetic field but depends on three
factors: a) the increase of the gravitational energy during the collapse, b)
the formation of a horizon, c) the reduction of the kinetic energy of
implosion. Such conditions are naturally met during the formation of an
extreme EMBH but are more general and can indeed occur in a variety of
different situations, e.g. during the formation of a Schwarzschild black hole
by a suitable extraction of the kinetic energy of implosion (see Fig.~\ref{fig1} and Fig.~\ref{fig3}).

Now consider a test particle of mass $m$ in the gravitational field of an already formed
Schwarzschild black hole of mass $M$ and go through such a sequence of braking
and energy extraction processes. Kaplan\cite{K49} found for the energy $E$ of
the particle as a function of the radius $r$
\begin{equation}
E=m\sqrt{1-\tfrac{2M}{r}}.\label{pointtest}%
\end{equation}
It would appear from this formula that the entire energy of a particle could
be extracted in the limit $r\rightarrow2M$. Such $100\%$ efficiency of energy
extraction has often been quoted as evidence for incompatibility between
General Relativity and the second principle of Thermodynamics (see Bekenstein
(1973)\cite{B73} and references therein). J. Bekenstein and S. Hawking have gone
as far as to consider General Relativity not to be a complete theory and to
conclude that in order to avoid inconsistencies with thermodynamics, the
theory should be implemented through a quantum description\cite{B73,H74}.
Einstein himself often expressed the opposite point of view (see.
e.g. Dyson (2002)\cite{D02} and references therein).

The analytic treatment presented in Cherubini, Ruffini, Vitagliano (2002)\cite{CRV02} can clarify this fundamental
issue. It allows to express the energy increase $E$ of a black hole of mass
$M_{1}$ through the accretion of a shell of mass $M_{0}$ starting its motion
at rest at a radius $R$ in the following formula which generalizes Eq.~(\ref{pointtest}):
\begin{equation}
E\equiv M-M_{1}=-\tfrac{M_{0}^{2}}{2R}+M_{0}\sqrt{1-\tfrac{2M_{1}}{R}},
\end{equation}
where $M=M_{1}+E$ is clearly the mass-energy of the final black hole. This
formula differs from the Kaplan formula (\ref{pointtest}) in three respects:
a) it takes into account the increase of the horizon area due to the accretion
of the shell; b) it shows the role of the gravitational self energy of the
imploding shell; c) it expresses the combined effects of a) and b) in an exact
closed formula.

The minimum value $E_{\mathrm{\min}}$ of $E$ is attained for the minimum value
of the radius $R=2M$: the horizon of the final black hole. This corresponds to
the maximum efficiency of the energy extraction. We have
\begin{equation}
E_{\min}=-\tfrac{M_{0}^{2}}{4M}+M_{0}\sqrt{1-\tfrac{M_{1}}{M}}=-\tfrac
{M_{0}^{2}}{4(M_{1}+E_{\min})}+M_{0}\sqrt{1-\tfrac{M_{1}}{M_{1}+E_{\min}}},
\end{equation}
or solving the quadratic equation and choosing the positive solution for
physical reasons
\begin{equation}
E_{\min}=\tfrac{1}{2}\left(  \sqrt{M_{1}^{2}+M_{0}^{2}}-M_{1}\right)  .
\end{equation}
The corresponding efficiency of energy extraction is
\begin{equation}
\eta_{\max}=\tfrac{M_{0}-E_{\min}}{M_{0}}=1-\tfrac{1}{2}\tfrac{M_{1}}{M_{0}%
}\left(  \sqrt{1+\tfrac{M_{0}^{2}}{M_{1}^{2}}}-1\right)  , \label{efficiency}%
\end{equation}
which is strictly \emph{smaller than} 100\% for \emph{any} given $M_{0}\neq0$.
It is interesting that this analytic formula, in the limit $M_{1}\ll M_{0}$,
properly reproduces the result of equation (\ref{Mirrmin}), corresponding to
an efficiency of $50\%$. In the opposite limit $M_{1}\gg M_{0}$ we have
\begin{equation}
\eta_{\max}\simeq1-\tfrac{1}{4}\tfrac{M_{0}}{M_{1}}.
\end{equation}
Only for $M_{0}\rightarrow0$, Eq.~(\ref{efficiency}) corresponds to an
efficiency of 100\% and correctly represents the limiting reversible
transformations introduced in Christodoulou, Ruffini (1971).\cite{CR71} It seems that the difficulties of
reconciling General Relativity and Thermodynamics are ascribable not to an
incompleteness of General Relativity but to the use of the Kaplan formula in a
regime in which it is not valid. The generalization of the above results to
stationary black holes is being considered.


\begin{thebibliography}{0}
\bibitem{CR71}D. Christodoulou, R. Ruffini, \emph{Phys. Rev. }\textbf{D4},
3552 (1971).

\bibitem{rlh} R. Ruffini, ``On the energetics of black holes'', in {\em Black Holes --- Les astres occlus}, edited by B. \& C. DeWitt, Gordon and Breach (New York, London, Paris, 1973).

\bibitem{r78}
R. Ruffini, ``Physics outside the horizon of a black hole'', in {\em Physics and Astrophysics of Neutron Stars and Black Holes}, Giacconi, R., Ruffini, R., Ed. and coauthors, North Holland, Amsterdam, 1978

\bibitem{r03}
Ruffini, R., Bianco, C.L., Chardonnet, P., Fraschetti, F., Xue, S.-S., 2003, IJMPD, submitted to

\bibitem{CRV02}C. Cherubini, R. Ruffini, L. Vitagliano, \emph{Phys. Lett. 
}\textbf{B545}, 226 (2002).

\bibitem{RV02}R. Ruffini, L. Vitagliano, \emph{Phys. Lett. }\textbf{B545},
233 (2002).

\bibitem{L32}L. Landau, \emph{Phys. Zeits. Sowj.} \textbf{1}, 285 (1932).

\bibitem{G51}G. Gamow, C.L. Critchfield, \emph{Theory of Atomic Nucleus and
Energy Sources} (Clarendon Press, Oxford, 1951).

\bibitem{OV39}J. R. Oppenheimer, G. Volkoff, \emph{Phys. Rev. } \textbf{D55},
374 (1939).

\bibitem{HTWW}B.K. Harrison, K.S. Thorne, M. Wakano, J.A. Wheeler,
\emph{Gravitation Theory and Gravitational Collapse} (University of
Chicago Press, Chicago, 1965).

\bibitem{GR75}H. Gursky, R. Ruffini, \emph{Neutron Stars, Black Holes and
X-Ray Sources} (Reidel, Dordrecht, 1975).

\bibitem{K49}S. A. Kaplan, \emph{Zh. Eksp. \& Teor. Fiz.} \textbf{19}, 951 (1949).

\bibitem{B73}J. Bekenstein, \emph{Phys. Rev. }\textbf{D7}, 2333 (1973).

\bibitem{H74}S. Hawking, \emph{Nature} \textbf{248}, 30 (1974).

\bibitem{D02}F. Dyson, \emph{communication at ``Science and Ultimate
Reality'', Symposium in honour of J. A. Wheeler,} Princeton\emph{ }(2002).
\end{thebibliography}
\end{document}